\documentstyle{elsart}
\newcommand{\decay}{$Z \rightarrow \nu \bar{\nu} \gamma$~}
\newcommand{\SULUY}{$ \mathrm{SU(2)_{L} \times U(1)_{Y}} $~}
\newcommand{\UEM}{$\mathrm{U(1)_{em}}~$}

\begin{document} 
\begin{frontmatter} 
\title{A Direct Constraint on Dimension--Eight Operators from \decay
\thanksref{CONA}}
\author[BUAP]{M. Maya},
\author[CINVES]{M. A. P\'erez \thanksref{emailmpa}},
\author[CINVES]{G. Tavares--Velasco \thanksref{emailgtv}} and
\author[BUAP]{B. Vega}
\address[BUAP]{Facultad de Ciencias F\'{\i}sico Matem\'aticas,
Universidad Auton\'oma de Puebla, Apartado Postal 1152, Puebla, M\'exico.}
\address[CINVES]{Departamento de F\'{\i}sica, CINVESTAV IPN,
\\Apartado Postal 14-740, 07000, M\'exico D. F.,M\'exico.}
\thanks[CONA]{This work was partially supported by CONACyT, M\'exico.} 
\thanks[emailmpa]{E--mail: mperez@fis.cinvestav.mx}
\thanks[emailgtv]{E--mail: gtv@fis.cinvestav.mx}

\begin{abstract} 

We study the radiative decay \decay within an effective Lagrangian
approach. Using the search for energetic single--photon events in the data
collected by the L3 Collaboration, we get direct bounds on dimension--six
and dimension--eight operators associated with the $\tau$--neutrino
magnetic moment and the anomalous electromagnetic proper\-ties of the $Z$
boson. As a by-product of our calculation, we reproduce the L3 result for
the bound on $\mu_{\nu_{\tau}}$. 

\end{abstract}

\end{frontmatter}

\maketitle

The effective Lagrangian approach concerning the local \SULUY symmetry
linearly realized \cite{Bur83}~ has been used recently to explore the
consequences of physics beyond the Standard Model (SM) at lepton
\cite{Hag97}, hadron \cite{Dia97}~and $\gamma \gamma$ colliders
\cite{Bel92}. Also, this approach has been used to constrain the anomalous
electromagnetic couplings of the $W$ boson, the $t$ quark \cite{Mar94}~
and the neutrinos \cite{Lar95}~ from the known experimental bounds on the
rare decays $b \rightarrow s \gamma$ \cite{Amm93} ~and $\mu \rightarrow
e\gamma$ \cite{PDG96}.  In the present letter we point out that the recent
measurement of energetic single--photons at LEP arising from the radiative
decay \decay \cite{Acc95}~ leads to direct constraints on dimension--eight
and dimension--six operators associated with the anomalous electromagnetic
properties of the $Z$ vector boson and the $\tau$--neutrino magnetic
moment, respectively. 

The radiative decay \decay can not be induced at the tree level in the SM. 
In the effective Lagrangian approach this decay could proceed through the
Feynman diagrams shown in Fig. \ref{fig1}, where the dots indicate
effective vertices induced by dimension--six or dimension--eight operators
which modify the SM weak sector. The anomalous $\nu \bar{\nu} \gamma$
vertex arises from the dimension--six operators \cite{Lar95}: 

  
\begin{figure} 
\input feynman 
\textheight800pt 
\textwidth 450pt
\begin{picture}(38000,42000) 
\drawline\photon[\E\REG](10000,33000)[8]
\put(6800,\photonfronty){$Z_{\mu}(p)$}
\drawline\fermion[\SE\REG](\photonbackx,\photonbacky)[8000]
\put(\fermionbackx,\fermionbacky){~~$\bar{\nu}(p_{1})$}
\put(\photonbackx,\fermionbacky){\makebox(0,0){\bf(a)}}
\drawarrow[\SE\ATTIP](\fermionbackx,\fermionbacky) 
\drawline\fermion[\NE\REG](\photonbackx,\photonbacky)[4000]
\drawline\fermion[\NE\REG](\fermionbackx,\fermionbacky)[4000]
\drawarrow[\NE\ATTIP](\fermionbackx,\fermionbacky) 
\put(\fermionbackx,\fermionbacky){~~$\nu(p_{2})$}
\drawline\photon[\E\REG](\fermionfrontx,\fermionfronty)[3]
\put(\fermionfrontx,\fermionfronty){\circle*{500}}
\put(\photonbackx,\photonbacky){~~$A_{\nu}(k)$}
\drawline\photon[\E\REG](10000,20000)[8]
\put(6800,\photonfronty){$Z_{\mu}(p)$}
\drawline\photon[\E\REG](\photonbackx,\photonbacky)[4]
\put(\photonbackx,\photonbacky){~~$A_{\nu}(k)$}
\drawline\fermion[\SE\REG](\photonfrontx,\photonfronty)[6000]
\put(\photonfrontx,\fermionbacky){\makebox(0,0){\bf(b)}}
\drawarrow[\SE\ATBASE](\fermionbackx,\fermionbacky) 
\put(\fermionbackx,\fermionbacky){~~$\bar{\nu}(p_{1})$}
\drawline\fermion[\NE\REG](\photonfrontx,\photonfronty)[6000]
\put(\fermionbackx,\fermionbacky){~~$\nu(p_{2})$}
\drawarrow[\NE\ATTIP](\fermionbackx,\fermionbacky) 
\put(\photonfrontx,\photonfronty){\circle*{500}}
\drawline\photon[\E\REG](10000,6500)[8]
\put(6800,\photonfronty){$Z_{\mu}(p)$}
\drawline\photon[\SE\REG](\photonbackx,\photonbacky)[3]
\drawline\fermion[\SE\REG](\photonbackx,\photonbacky)[4000]
\put(\fermionbackx,\fermionbacky){~~$\bar{\nu}(p_{1})$}
\drawarrow[\SE\ATTIP](\fermionbackx,\fermionbacky) 
\drawline\fermion[\NE\REG](\photonbackx,\photonbacky)[4000]
\drawarrow[\NE\ATTIP](\fermionbackx,\fermionbacky) 
\put(\fermionbackx,\fermionbacky){~~$\nu(p_{2})$}
\drawline\photon[\NE\REG](\photonfrontx,\photonfronty)[6]
\put(\photonbackx,\photonbacky){~$A_{\nu}(k)$}
\put(\photonfrontx,\photonfronty){\circle*{500}}
\global\advance\photonfronty by -5000
\put(\photonfrontx,\photonfronty){\makebox(0,0){\bf(c)}}
\end{picture} 
\caption{Feynman diagrams
contributing to the decay \decay in the
effective Lagrangian approach.  The heavy dots
denote effective vertices.} \label{fig1}
\end{figure}
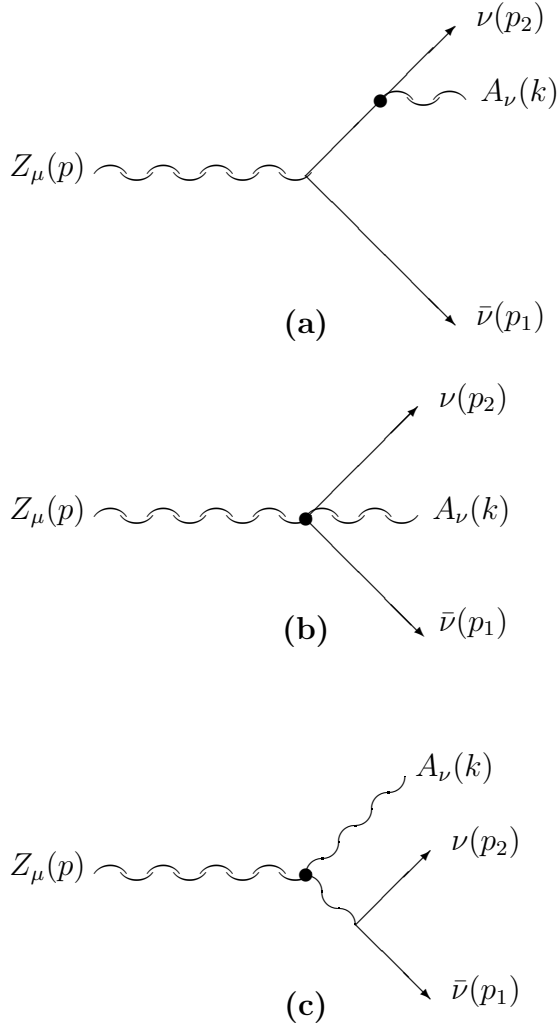 \vskip 0.3 cm

\begin{equation} 
O^{ab}_{\nu W}=\bar{\ell}^{a}_{L} \sigma^{\mu \nu}
W^i_{\mu \nu} \tau^{i} \tilde{\phi}
\nu^{b}_{R}, \label{sixop1} 
\end{equation}

\begin{equation} 
O^{ab}_{\nu
B}=\bar{\ell}^{a}_{L} \sigma^{\mu \nu} B_{\mu
\nu} \tilde{\phi} \nu^{b}_{R}, \label{sixop2}
\end{equation}

while the four--point vertex $Z \nu \bar{\nu}
\gamma$ arises from the dimension--eight
operators

\begin{equation} 
O^{8}_{1}={\mathrm{i}}
(\phi^{\dagger} \phi)\bar{\ell}^{a}_{L}
\tau^{i} \gamma^{\mu}
D^{\nu}\ell^{a}_{L}W^{i}_{ \mu \nu},
\label{eiop1} 
\end{equation}

\begin{equation} 
O^{8}_{2}= {\mathrm{i}}
(\phi^{\dagger} \phi)\bar{\ell}^{a}_{L}
\gamma^{\mu} D^{\nu} \ell^{a}_{L} B_{\mu \nu},
\label{eiop2} 
\end{equation}

\begin{equation}
O^{8}_{3}= {\mathrm{i}} (\phi^{\dagger}
D^{\mu} \phi)\bar{\ell}^{a}_{L} \gamma^{\nu}
\tau^{i} \ell^{a}_{L}W^{i}_{\mu \nu},
\label{eiop3}
\end{equation}

\begin{equation}
O^{8}_{4}= {\mathrm{i}} (\phi^{\dagger} D ^{\mu}
\phi)\bar{\ell}^{a}_{L}
\gamma^{\nu}
\ell^{a}_{L} B_{\mu \nu}. \label{eiop4}
\end{equation}

All these operators preserve the \SULUY SM gauge simmetry. We have denoted
with standard notation the $\mathrm{SU(2)_{L}}$ and $ \mathrm{U(1)_{Y}}$
tensor field strength tensors $W^{i}_{\mu \nu}$ and $B_{\mu \nu}$,
respectively, as well as the $ \mathrm{SU(2)_{L}}$ left--handed lepton
doublet $\ell^{a}_{L}$, the respective right--handed neutrinos
$\nu^{a}_{R}$, the Pauli matrices $\tau_{i}$, the Higgs field
$\tilde{\phi}=\mathrm{i} \tau^{2} \phi^{\ast}$ and the respective
covariant derivative $D_{\mu}$ \cite{Wud94}. 

Previous studies on the anomalous $ZZ \gamma$ coupling shown in Fig.
\ref{fig1}(c) have used an \UEM gauge invariant parametrization
\cite{Acc95,Bau93}. However, it is important to notice that in the
effective Lagrangian approach there are no effective operators of
dimension lower than eight, which are \SULUY gauge invariant and could
lead to an anomalous $ZZ \gamma$ vertex. As a consequence, it is expected
that this vertex is highly suppressed in the SM. This situation has been
confirmed by an \UEM gauge invariant calculation of the CP-conserving and
CP-violating off-shell $ZZ \gamma$ vertex \cite{Bar95}. Since in the
effective Lagrangian approach the non--standard $ZZ \gamma$ vertex can not
be generated at this level, we will not consider its effect on the \decay
radiative decay. 

It is possible to establish in the effective Lagrangian approach the order
of perturbation theory in which SM gauge invariant non--renormalizable
operators may be generated in the underlying theory\cite{Art95}. In
particular, loop generated operators appear with a characteristic
suppression factor $\sim 1/(4 \pi)^2$ which significantly decreases the
magnitude of their effects. For example, in the case of the radiative
decay of the SM Higgs boson into two photons, it was found that some tree
level generated operators of dimension--eight may compete with
dimension--six operators which are generated at the one--loop
level\cite{Her95}. In our case, the dimension--six operators
(\ref{sixop1})--(\ref{sixop2}) are induced at the one--loop level in the
underlying theory, whereas the dimension--eight ope\-rators
(\ref{eiop1})-(\ref{eiop4}) are induced at the tree level. As a
consequence, we expect that all these ope\-rators give similar
contributions to the \decay decay trough the anomalous $\nu \bar{\nu}
\gamma$ and $Z \nu \bar{\nu} \gamma$ vertices. We will ignore
CP--violating effects in this decay relying on general expectations that
the scale of CP--violation is greater than $\Lambda$, the scale used in
the effective Lagrangian approach to denote the characteristic energy in
which non--standard effects are expected to become directly observable. 

After spontaneous symmetry breaking, the operators
(\ref{sixop1})-(\ref{eiop4}) induce the fo\-llo\-wing parametrization for
the $\nu \bar{\nu} \gamma$ and $Z \nu \bar{\nu} \gamma$ effective
couplings,

\begin{equation}
M^{(a)}_{\mu \nu}=\frac{\epsilon_6}{v}
\bar{u}(p_2)\gamma_{\mu}
(g_{\mathrm{V}}-g_{\mathrm{A}}\gamma_5)\FMSlash{q}^{-1}\sigma_{\nu
\alpha} k^{\alpha} v(p_1), \label{EcMa}
\end{equation}

\begin{equation} M^{(b)}_{\mu
\nu}=\frac{\epsilon_8}{v^2}\bar{u}(p_2)(1-\gamma_5)(k_{\mu}
\gamma_{\nu}-\FMSlash{k}g_{\mu \nu}) v(p_1),
\label{EcMb} \end{equation}

where we have used the kinematic variables shown in Fig.\ref{fig1},
$q=k+p_{2}$, $g_{\mathrm{V,A}}=g/{4 c_{w}}$ are the couplings of $Z$ to
neutrinos in the SM and $v$ is the SM vacuum expectation value.  The
coefficients $\epsilon_{6,8}$ summarize all the information that can be
gathered from the heavy degrees of freedom associated with new physics
effects and are expressed in terms of dimensionless coupling constants
$\alpha_{i}$ and the scale $\Lambda$:  $\epsilon_{6}=\epsilon_{6}^{\nu W}
+ \epsilon_{8}^{\nu B}$ and $\epsilon_{8}=\epsilon_{8}^{2} +
\epsilon_{8}^{3} + \epsilon_{8}^{4}$; with
$\epsilon^{i}_{6}=\alpha^{i}_{6}(v/\Lambda)^2$ and
$\epsilon^{i}_{8}=\alpha^{i}_{8}(v/\Lambda)^4$.

It is easy to see that contributions (\ref{EcMa}) and (\ref{EcMb}) do not
interfere. Furthermore, assuming the simple situation that cancellation
among different operators does not take place, we get the following
expressions for the distribution of the photon energy ($x=E_{k}/M_Z$)
arising from diagrams \ref{fig1}(a)  and \ref{fig1}(b),

\begin{equation}
\frac{\d\Gamma^{(a)}}{\d
x}=\frac{\epsilon^{2}_{6}
(g^2_{\mathrm{V}}+g^2_{\mathrm{A}}) M^3_Z}{72
\pi^3 v^2} x(3(1-2x)+x^2), \label{Disa}
\end{equation}

\begin{equation}
\frac{\d\Gamma^{(b)}}{\d
x}=\frac{\epsilon^2_8 M^5_Z}{18
\pi^3 v^4} x^3(1-x). \label{Disb}
\end{equation}

The measurement of energetic single--photons at LEP arising from the decay
\decay has been used to set a direct limit on the $ZZ \gamma$ \UEM--gauge
invariant coupling and the magnetic moment of the $\tau$ neutrino. For the
purposes of the present analysis in the framework of the effective
Lagrangian approach, the search for the energetic single--photons events
on the data collected by the L3 collaboration may be translated easily
into bounds on the coefficients $\epsilon_6$ and $\epsilon_8$ contained in
the energy distributions (\ref{Disa}) and (\ref{Disb}). In order to reduce
backgrounds, the L3 collaboration required the photon energy to be greater
than one half the $e^+e^-$ beam energy. The L3 collaboration obtained a
branching ratio limit of one part in a million when the energy of the
photon in \decay is above 30 Gev. Integrating (\ref{Disa}) and
(\ref{Disb}) over the relevant range of energy we obtain the following
bounds on the $\epsilon_{6,8}$ coefficients which correspond to the two
events selected from the L3 data

\begin{equation}
\epsilon_6 < 0.192, \label{lime6}
\end{equation}

\begin{equation}
\epsilon_8 < 0.165 . \label{lime8}
\end{equation}

The constraint (\ref{lime6}) can be translated into an upper limit on the
$\tau$--neutrino magnetic moment in units of Bohr magnetons
$\mu_{\mathrm{B}}$,

\begin{equation} 
\mu_{\nu_{\tau}}
<\mathrm{2.62} \times \mathrm{10^{-6}}
~\mathrm{\mu_B}.  \label{limmagmom}
\end{equation}

Our bound (\ref{limmagmom}) is consistent with the L3 limit
$\mu_{\nu_{\tau}}< 3 \times10^{-6}$ \cite{Acc95}. This means that the
determination of this quantity is independent of the scale involved in the
effective vertex (\ref{EcMa}):  the dimension--six operators
(\ref{sixop1})  and (\ref{sixop2}) induce in our case a scale given by
$v$, while the L3 limit used the traditional scale given in terms of the
electron mass. Maltoni and Vysotski \cite{Mal98} reproduced our
calculation for $\Gamma^{(a)}$ and $\Gamma^{(b)}$ recently.  Our bounds
given in (\ref{lime6})-(\ref{limmagmom}) agree with their results on the
coefficients $\epsilon_{6,8}$ and $\mu_{\nu_{\tau}}$. There is a small
difference among their bounds and ours due to the fact that we are
consi\-de\-ring the L3 branching ratio and the known experimental value
for the full $Z$ width decay \cite{PDG96}, while they considered directly
the L3 value for $N_{Z\rightarrow\mathrm{had}}$. In particular, our bound
on $\mu_{\nu_{\tau}}$ compares favourably with the bound
$\mu_{\nu_{\tau}}< \mathrm{4} \times \mathrm{10^{-6}} ~\mathrm{\mu_B}$
obtained from low--energy expe\-riments \cite{Gro88} and
$\mu_{\nu_{\tau}}< \mathrm{2.7} \times \mathrm{10^{-6}} ~\mathrm{\mu_B}$
obtained from the invisible width of the $Z$ boson \cite{Esc92}, and it is
close to the one derived from a beam--dump experiment \cite{Coo92}. It is
interestig to notice that these bounds on the $\tau$--neutrino magnetic
moment are still weaker than the known bounds on the magnetic moments of
electron and muon neutrinos \cite{Kra90} and the transition magnetic
moments $\nu_{\tau} \rightarrow\nu_{i}\gamma$ obtained from the
experimental bound on $\mu \rightarrow e \gamma$ also within the effective
Lagrangian approach~\cite{Lar95}. 

In conclusion, we have obtained direct bounds on the $\tau$--neutrino
magnetic moment and the dimension--eight operators
(\ref{eiop1})-(\ref{eiop4}). These bounds reflect the natural consequence
that non--standard effects may become enhanced when SM contributions are
highly suppressed. We know that this is the situation with the magnetic
moments of the neutrinos and the $ZZ \gamma$ vertex. In the case of the $Z
\nu \bar{\nu} \gamma$ effective vertex, the SM calculation for the
respective box diagrams is not available yet, but we expect to have a
similar situation in this case \cite{Per97}.

\ack{We thank important suggestions made by J.  M. Hern\'andez, F. Larios
and J. J. Toscano.}

%


\begin{thebibliography}{9}

\bibitem{Bur83} C.J.C. Burguess and H.J. Schnitzer, Nucl. Phys. B228
(1983) 464; C.N. 
Leungy, S.J. Love and S. Rao, Z. Phys. 31 (1986) 433; W. Buchm\"uller and 
D.
Wyler, 
Nucl. Phys. B268 (1986) 621; K. Hagiwara, H. Hikasa, R.D. Peccei and D. 
Zeppenfeld, Nucl. Phys. B 282 (1987) 253.

\bibitem{Hag97} K. Hagiwara, T. Hatsukano, S. Ishikara and R. Szalapski,
Nucl. Phys. 
B496 (1997) 66; K. Whisnant, J.M. Yang, B.-L. Young and X. Zhang, Phys.
Rev. D56 
(1997) 467; G.J. Gounaris, D.T. Papadamon and F.M. Renard, Z. Phys. 76
(1997) 
333; A. Datta, K. Shiwnant, Bol. Young and Z. Zhang, Phys. Rev. D57
(1998)364.

\bibitem{Dia97} J.L. D\'{\i}az-Cruz, M.A. P\'erez and J.J. Toscano, Phys.
Lett. B398 
(1997) 347; F. de Campos, M.C. Gonz\'alez-Garc\'{\i}a and S.F. Novaes,
Phys. Rev. 
Lett. 769 (1997) 5210.

\bibitem{Bel92} G. B\'elanger and F. Boudjema, Phys. Lett. B288 (1992) 21;
O.J.P. 
Eboli et al., Phys. Rev. D52 (1995) 15; J.M. Hern\'andez, M.A. P\'erez and 
J.J. 
Toscano, Phys. Lett. B375 (1996) 227.

\bibitem{Mar94} R. Mart\'{\i}nez, M.A. P\'erez and J.J. Toscano, Phys.
Lett. B340 
(1994); T.G. Rizzo, Phys. Lett. B315(1993); X. He and B. Mc
Kellar, Phys. Lett. B320 (1994) 165.

\bibitem{Lar95} F. Larios, R. Mart\'{\i}nez and  M.A. P\'erez, Phys. Lett.
B345 (1995) 
259.

\bibitem{Amm93} R. Ammar et al., CLEO Collab. Phys. Rev. Lett. 71 (1993)
674.

\bibitem{PDG96} Particle Data Group, R. Barnett {\it et al.} Review of particle 
properties, Phys. Rev. D54 (1996) 1.

\bibitem{Acc95} M. Acciarri {\it et al}., L3 Collab., Phys.
 Lett. B412 (1997) 201; 
Phys. Lett. B346 (1995) 190. 

\bibitem{Wud94}J. Wudka, Int. J. Mod. Phys. A9 (1994) 2301;
J. Feliciano et al., Rev. Mex. F\'{\i}s. 42 ~~(1996) 517.

\bibitem{Bau93} U. Baur and E.L. Berger, Phys. Rev. D47
(1993); D. Choudhury and ~~S.D. Rindani, Phys. Lett. B335 (1994) 198. 

\bibitem{Bar95} A. Barroso, F. Boudjema, J. Cole and N. Dombey, Z. Phys.
C28 (1995) 
149.

\bibitem{Art95} C. Artz, M.B. Einhorn and J. Wudka, Nucl. Phys. B433
(1995) 41.

\bibitem{Her95} J.M. Hern\'andez, M.A. P\'erez and J.J. Toscano, Phys.
Rev. D51 
(1995) 2044.
\bibitem{Mal98} M. Maltoni and M. I. Vysotsky, hep-ph/9804464 preprint
(unpublished).

\bibitem{Gro88} H. Grotch and R. Robinett, Z. Phys. C39 (1988) 553.

\bibitem{Esc92} R. Escribano and E. Masso, Phys. Lett. B280 (1992) 153.

\bibitem{Coo92} A.M. Cooper-Sarkar et al., Phys. Lett. B280 (1992) 153.

\bibitem{Kra90} X. Krakaner {\it et al}., Phys. Lett. B252 (1990) 177.

\bibitem{Per97} M.A. P\'erez, G. Tavares--Velasco and J.J. Toscano, work
in
progress.

\end{thebibliography}
\end{document}